\documentclass[preprint,showpacs,preprintnumbers,amsmath,amssymb]{revtex4}

\usepackage{graphicx}
\usepackage{dcolumn}
\usepackage{bm}
\usepackage{epsf,epsfig,latexsym}

\newcommand{\Version}{3 December 2001 }

\newif\ifUseBibTeX\UseBibTeXtrue 
\UseBibTeXfalse 

\newif\ifRevTexAuthor\RevTexAuthortrue
\RevTexAuthorfalse

\ifRevTexAuthor
\newcommand{\bnl}{Brookhaven National Laboratory, Upton,New York
  11973}
\newcommand{\ires}{Institut de Recherches Subatomiques and
  Universit{\'e} Louis Pasteur,Strasbourg, France}
\newcommand{\kraknuc}{Institute of Nuclear Physics, Krakow, Poland}
\newcommand{\krakow}{Jagiellonian University, Krakow, Poland}
\newcommand{\baltimore}{Johns Hopkins University, Baltimore, Maryland
  21218}
\newcommand{\newyork}{New York University, New York, New York 10003}
\newcommand{\nbi}{Niels Bohr Institute, University of Copenhagen,
  Denmark}
\newcommand{\texas}{Texas A$\&$M University, College Station,Texas
  77843}
\newcommand{\bergen}{University of Bergen, Department of Physics,
  Bergen,Norway}
\newcommand{\bucharest}{University of Bucharest,Romania}
\newcommand{\kansas}{University of Kansas, Lawrence, Kansas 66045}
\newcommand{\oslo}{University of Oslo, Department of Physics,
  Oslo,Norway}
\else 
\newcommand{\bnl}           {$\rm^{1}$}
\newcommand{\ires}          {$\rm^{2}$}
\newcommand{\kraknuc}       {$\rm^{3}$}
\newcommand{\krakow}        {$\rm^{4}$}
\newcommand{\baltimore}     {$\rm^{5}$}
\newcommand{\newyork}       {$\rm^{6}$}
\newcommand{\nbi}           {$\rm^{7}$}
\newcommand{\texas}         {$\rm^{8}$}
\newcommand{\bergen}        {$\rm^{9}$}
\newcommand{\bucharest}     {$\rm^{10}$}
\newcommand{\kansas}        {$\rm^{11}$}
\newcommand{\oslo}          {$\rm^{12}$}
\fi

\begin{document}

\title{Pseudorapidity distributions of charged particles from  Au+Au
  collisions at the maximum RHIC energy, $\sqrt{s_{NN}}$=200~GeV} 
\ifRevTexAuthor
\else 
\author{
  I.~G.~Bearden\nbi, 
  D.~Beavis\bnl, 
  C.~Besliu\bucharest, 
  Y.~Blyakhman\newyork, 
  B.~Budick\newyork, 
  H.~B{\o}ggild\nbi, 
  C.~Chasman\bnl, 
  C.~H.~Christensen\nbi, 
  P.~Christiansen\nbi, 
  J.~Cibor\kraknuc, 
  R.~Debbe\bnl, 
  E. Enger\oslo, 
  J.~J.~Gaardh{\o}je\nbi, 
  K.~Hagel\texas, 
  O.~Hansen\nbi, 
  A.~Holm\nbi, 
  A.~K.~Holme\oslo, 
  H.~Ito\kansas, 
  E.~Jakobsen\nbi, 
  A.~Jipa\bucharest, 
  J.~I.~J{\o}rdre\bergen, 
  F.~Jundt\ires, 
  C.~E.~J{\o}rgensen\nbi, 
  R.~Karabowicz\krakow, 
  T.~Keutgen\texas, 
  E.~J.~Kim\bnl, 
  T.~Kozik\krakow, 
  T.~M.~Larsen\oslo, 
  J.~H.~Lee\bnl, 
  Y.~K.~Lee\baltimore, 
  G.~L{\o}vh{\o}iden\oslo, 
  Z.~Majka\krakow, 
  A.~Makeev\texas, 
  B.~McBreen\bnl, 
  M.~Mikelsen\oslo, 
  M.~Murray\texas, 
  J.~Natowitz\texas, 
  B.~S.~Nielsen\nbi, 
  J.~Norris\kansas, 
  K.~Olchanski\bnl, 
  J.~Olness\bnl, 
  D.~Ouerdane\nbi, 
  R.~P\l aneta\krakow, 
  F.~Rami\ires, 
  C.~Ristea\bucharest, 
  D.~R{\"o}hrich\bergen, 
  B.~H.~Samset\oslo, 
  D.~Sandberg\nbi, 
  S.~J.~Sanders\kansas, 
  R.~A.~Sheetz\bnl, 
  P.~Staszel\nbi, 
  T.~F.~Thorsteinsen\bergen$^+$,
  T.~S.~Tveter\oslo, 
  F.~Videb{\ae}k\bnl, 
  R.~Wada\texas, 
  A.~Wieloch\krakow, and
  I.~S.~Zgura\bucharest\\
  (BRAHMS Collaboration )\\[1ex]
  \bnl~Brookhaven National Laboratory, Upton,New York 11973,
  \ires~Institut de Recherches Subatomiques and Universit{\'e} Louis
  Pasteur, Strasbourg, France,
  \kraknuc~Institute of Nuclear Physics, Krakow, Poland,
  \krakow~Jagiellonian University, Krakow, Poland,
  \baltimore~Johns Hopkins University, Baltimore, Maryland 21218,
  \newyork~New York University, New York, New York 10003,
  \nbi~Niels Bohr Institute, University of Copenhagen, Denmark,
  \texas~Texas A$\&$M University, College Station,Texas 77843,
  \bergen~University of Bergen, Department of Physics, Bergen,Norway,
  \bucharest~University of Bucharest,Romania,
  \kansas~University of Kansas, Lawrence, Kansas 66045,
  \oslo~University of Oslo, Department of Physics, Oslo,Norway,
  $^+ Deceased$}
\noaffiliation
\fi
\date{\Version}

\begin{abstract}
  We present charged-particle multiplicities as a function of
  pseudorapidity and collision centrality for the
  $^{197}$Au+$^{197}$Au reaction at $\sqrt{s_{NN}}$=200~GeV.  For the
  5\% most central events we obtain 
  $dN_{ch}/d\eta|_{\eta = 0}$ = 625$\pm$55
  and 
  $N_{ch} |_{-4.7\le \eta \le 4.7} =  4630\pm 370$,  i.e.
  14\% and 21\% increases, respectively,  
  relative to $\sqrt{s_{NN}}$=130~GeV collisions.  Charged-particle 
  production per pair of participant nucleons is found
  to increase from peripheral 
  to central collisions around mid-rapidity.  These results
  constrain current models of particle production at the highest
  RHIC energy.
\end{abstract}
\maketitle

A central question in the study of collisions between heavy nuclei at
the maximum energy of the RHIC facility,
$\sqrt{s_{NN}}$=200~GeV, is the role of hard scatterings between
partons and the interactions of these partons in a high-density
environment. A reduction in the number of hadrons at large
transverse momentum has 
already been observed for $\sqrt{s_{NN}}$=130~GeV collisions 
that may hint at suppression of hadronic jets at high matter
densities~\cite{Phenix-jets,Star-jets}.
More generally, it has been conjectured that new phenomena
related to non-perturbative QCD may occur at the
highest RHIC energy. For example, a 
saturation of the number of parton
collisions in central nucleus-nucleus collisions could lead to a
limit on the  production of charged
particles~\cite{partonsat83,Eskola00,Kharzeev_and_Levin}.  

The present Letter addresses these issues with the first comprehensive
investigation of multiplicity distributions of emitted charged
particles in relativistic collisions between $^{197}$Au
nuclei with $\sqrt{s_{NN}}$=200~GeV. In particular, we have measured
pseudorapidity distributions of charged particles 
$dN_{ch}/d\eta$ in the range $-4.7 \le \eta \le 4.7$  
as a function of collision
centrality.  The pseudorapidity variable $\eta$ is related to the particle
emission angle $\theta$ with $\eta=-{\rm ln}[{\rm tan}(\theta/2)]$.
We find that the
production of charged particles at mid-rapidity ($\eta \approx 0$) 
increases by  (14$\pm$4)\%
for the most central collisions relative to $\sqrt{s_{NN}}$=130~GeV
collisions~\cite{back00,Star-mult130,adcox01,bearden01a}, in agreement 
with results of the PHOBOS experiment~\cite{Phobos-mult200-1}.
In highly
energetic nuclear collisions, charged particles can 
be produced by hadronic (``soft'') as well as
partonic (``hard'') collision processes. By extending the 
$dN_{ch}/d\eta$ systematics to cover a range of reaction centralities
and pseudorapidities, it becomes possible to more fully explore the
different reaction mechanisms.
  
The data were obtained using several subsystems of the  
BRAHMS experiment at RHIC~\cite{bearden01b}: 
the Multiplicity Array (MA), the
Beam-Beam Counter (BBC) arrays, and the Zero-Degree Calorimeters
(ZDCs).  An analysis of
charged-particle multiplicities for Au+Au reactions at
$\sqrt{s_{NN}}$=130~GeV using a nearly identical procedure 
to the one presented here is described in ref.~\cite{bearden01a}. 

The MA determines $dN_{ch}/d\eta$ around mid-rapidity 
with a modestly segmented Si-strip-detector array (SiMA) 
surrounded by an outer
plastic-scintillator tile array (TMA) in 
a double, hexagonal-sided barrel arrangement.   
Each of the 25 Si detectors (4~cm x 6~cm x 300~$\mu$m) is located 5.3~cm from
the beam axis and is subdivided along the beam direction
into seven active strips. The TMA 
has 35 plastic-scintillator tiles (12~cm x 12~cm x 0.5~cm) 
located 13.9 cm from the beam axis.  The effective coverage of the MA is
$-3.0\le\eta\le3.0$. 
The SiMA 
is used alone for determining $dN_{ch}/d\eta$ values near mid-rapidity
because of its higher segmentation.
However, both the SiMA and TMA are
used for establishing reaction centrality, as discussed below.  
Particle multiplicities are deduced from the observed
energy loss in the SiMA and TMA
elements by using GEANT simulations~\cite{Geant} 
to relate energy loss to the number of particles hitting
a given detector element~\cite{bearden01a}. 
SiMA and TMA elements are calibrated using
low-multiplicity events where well-defined peaks are observed
in the individual energy spectra corresponding to
single-particle hits~\cite{bearden01a}. 

The BBC arrays consist of two sets of Cherenkov UV-transmitting
plastic radiators coupled to photomultiplier tubes. The
Cherenkov radiators are
positioned around the beam pipe with one set 
on either side of the nominal
interaction point at a distance of 2.20~m.  The time resolution of the
BBC elements permits the determination of the interaction point with an
accuracy of $\approx$ 0.9~cm.  Charged-particle multiplicities with 
$2.1\le |\eta| \le 4.7$ are deduced from the
number of particles hitting each detector, as found by dividing the
measured detector signal by the average response of the detector
to a single particle.

The ZDCs are located $\pm$18~m from the nominal
interaction point and measure neutrons that are emitted at small
angles with respect to the beam direction~\cite{adler00}.  Clean
selection of minimum-biased events required a coincidence between the two
ZDC detectors and a minimum of 4 ``hits'' in the TMA. It is estimated that
this selection 
includes 95\% of the Au+Au total inelastic cross section. 

\begin{figure}
  \epsfig{file=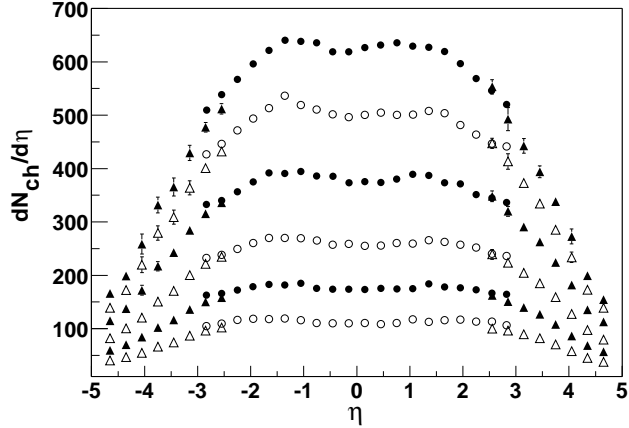,width=8.5cm}
  \caption{
    Distributions of $dN_{ch}/d\eta$ for centrality ranges of, top to
    bottom, 0-5\%, 5-10\%, 10-20\%, 20-30\%, 30-40\%, and 40-50\%.
    The SiMA and BBC results are indicated by circles and
    triangles, respectively. Statistical errors are shown for all
    points where they are larger than the symbol size.}
  \label{dndeta}
\end{figure}

Reaction centrality is determined by selecting different regions 
in the total multiplicity distribution of either the MA or BBC arrays. 
The distributions are adjusted for ``missed'' events as
described in ref.~\cite{bearden01a}.
In determining $dN_{ch}/d\eta$, the centrality dependence 
of the MA and BBC distributions are based on the total multiplicity  
measurements of the corresponding array, thus allowing 
a range of vertex locations to be used in the BBC analysis 
beyond the acceptance of the MA (see ref.~\cite{bearden01a}). 
For
3.0$\le|\eta|\le$4.2, where it was possible to analyze the BBC data
using both centrality selections, the two analyses give
results to within 2\% of each other. In general, statistical errors on
the measurements are less than 1\%, with 
systematic errors of 8\% and 10\% for the SiMA and BBC arrays,
respectively. The systematic errors are dominated by overall scaling 
uncertainties resulting from the calibration procedures 
and should primarily lead  to a common
scale offset for data obtained at the two RHIC energies. 
However, there may
be as much as a 3\% relative scale error between the two energies. 
A point-to-point error is 
also present, as indicated by the small asymmetry seen in Fig.~\ref{dndeta} 
for the more central collisions. 

Figure~\ref{dndeta} shows the measured $dN_{ch}/d\eta$
distributions for charged particles for several centrality ranges.  
The
$dN_{ch}/d\eta$ values for these selected centralities 
at $\eta$=0, 3.0, 4.5 are listed
in Table~\ref{tb:dndeta}, together with the average number of participating
nucleons $\langle N_{part}\rangle$ 
estimated from the HIJING (heavy-ion jet interaction generator) model~\cite{wang91} 
using default parameters.  For the most central collisions (0-5\%), 
$dN_{ch}/d\eta |_{\eta = 0}=625\pm 1(stat)\pm 55(syst)$. 
This gives a scaled multiplicity value of
$(dN_{ch}/d\eta)/\langle N_{part}/2\rangle= 
3.5 \pm 0.3$ charged particles per
participating nucleon 
pair and indicates a
$(13\pm 4)\%$ increase 
relative to Au+Au reactions at
$\sqrt{s_{NN}}$=130~GeV
~\cite{bearden01a,sionly}. 
For
the most peripheral collisions analyzed here (40-50\%),
we find
$dN_{ch}/d\eta |_{\eta =0}=110\pm 10$, resulting in a 
scaled value of 3.0$\pm 0.4$. 
By integrating the 0-5\% multiplicity distribution we deduce that $4630
\pm 370$ charged particles are emitted in the considered pseudorapidity
range.  This value is $(21\pm 4)\%$ higher than at
$\sqrt{s_{NN}}$=130~GeV~\cite{bearden01a}.   

\begin{table} [ht]
  \caption{\label{tb:dndeta} 
    Table of $dN_{ch}/d\eta$ values. 
    Total uncertainties, dominated by
    the systematics, are indicated. The average
    number of participants $\langle N_{part}\rangle$ and collisions
    $\langle N_{coll}\rangle$ is given for each centrality
    class, with combined model and experimental uncertainties. The model
    uncertainties are obtained by varying the assumptions for the Glauber
    picture, including the NN cross section and the nuclear radius and
    surface diffuseness parameters. 
    $N_{ch}$ is the
    integral charged-particle multiplicity within 
    $-4.7 \le \eta \le 4.7$.}
  \begin{tabular}{ccccccc}
    \hline
    Cent- &
$\eta = 0$ &
$\eta = 3.0$ &
$\eta = 4.5$ &
$N_{ch}$ &
$\langle N_{coll}\rangle$&
$\langle N_{part}\rangle$
    \\
    rality &  & & & & & \\
    0-5 & 625$\pm$55 & 470$\pm$44
    & 181 $\pm$22 & 4630$\pm$370& 1000$\pm$125 & 357$\pm$8 \\
    5-10 & 501$\pm$44 & 397$\pm$37
    & 156$\pm$18 & 3810$\pm$300& 785$\pm$115 & 306$\pm$11 \\
    10-20 & 377$\pm$33 & 309$\pm$28
    & 125$\pm$14 & 2920$\pm$230 & 552$\pm$100 & 239$\pm$10 \\
    20-30 & 257$\pm$23 & 216$\pm$17
    & 90 $\pm$10 & 2020$\pm$160 & 335$\pm$58 & 168$\pm$9 \\
    30-40 & 174$\pm$16 & 149$\pm$14
    & 64 $\pm$7 & 1380$\pm$110 & 192$\pm$43 & 114$\pm$9 \\
    40-50 & 110$\pm$10 & 95$\pm$9
    & 43 $\pm$5 & 890$\pm$70 & 103$\pm$31 & 73$\pm$8 \\
    \hline
  \end{tabular}
\end{table}

\begin{figure}
  \epsfig{file=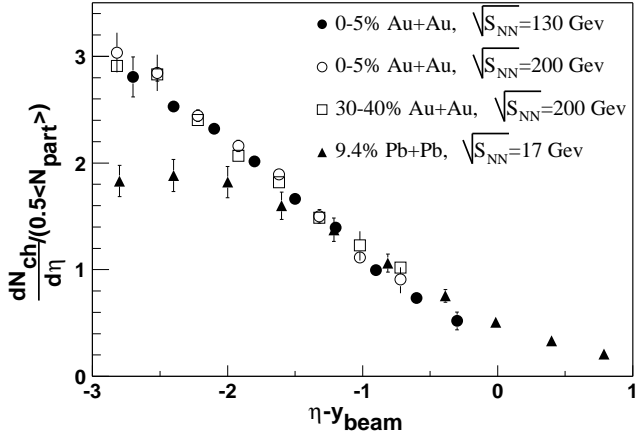,width=8.5cm}
  \caption{
    Charged-particle multiplicities normalized to the number of participant
    nucleon pairs (see Table I) 
    for the present 0-5\% central (open circles) and
    40-50\% central (open squares) Au+Au results at 
    $\sqrt{s_{NN}}$=200~GeV, the BRAHMS 0-5\% Au+Au
    results~\cite{bearden01a} at $\sqrt{s_{NN}}$=130~GeV (closed
    circles) and the 9.4\% central Pb+Pb data at
    $\sqrt{s_{NN}}$=17~GeV(closed triangles) of ref~\cite{deines00}.
    Data are plotted as a function of the
    pseudorapidity shifted by the relevant beam rapidity. Representative total
    uncertainties are shown for a few Au+Au points. }
  \label{dndeta_fragment}
\end{figure}

While the scaled multiplicities increase with centrality at
mid-rapidity, Fig.~\ref{dndeta_fragment} shows 
they are independent of both collision centrality and
beam energy over a pseudorapidity range from  0.5 to 1.5 units
below the beam rapidity. This is found for energies
ranging from the CERN-SPS energy 
($\sqrt{s_{NN}}$=17~GeV)
\cite{deines00} to the present RHIC beam energy and is consistent with a
limiting-fragmentation picture in which the excitations of the
fragment baryons saturate at a moderate collision energy,
independent of system size~\cite{bearden01a}. The
increased projectile kinetic energy is utilized for particle
production below beam rapidity, as evidenced by the
observed increase in the scaled multiplicity for central events
at mid-rapidity.  

\begin{figure}
  \epsfig{file=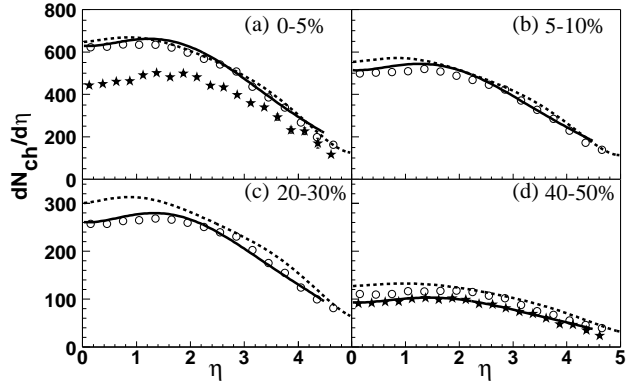,width=8.5cm}
  \caption{
    (a-d) Measured $dN_{ch}/d\eta$ distributions for centrality
    ranges of  0-5\%, 5-10\%, 20-30\% and 40-50\%. Theoretical
    predictions by Kharzeev and Levin (solid line) and by the
    AMPT model (dashed line) are also shown. Result from $p\bar p$ 
    collisions 
    at $\sqrt{s}$=200~GeV~\cite{Alner86}, scaled by the Au+Au
    values of $\langle N_{part}\rangle/2$, 
    are shown with stars (a,d).}
  \label{dndeta_models}
\end{figure}

Figure~\ref{dndeta_models} presents the $dN_{ch}/d\eta$ distributions
obtained by averaging the values for negative and positive 
pseudorapidities to further decrease the
experimental uncertainties. The solid
lines are calculations using the model of Kharzeev and
Levin~\cite{Kharzeev_and_Levin}. This model, which is based on a 
classical QCD
calculation using parameters fixed to the $\sqrt{s_{NN}}$=130~GeV
data, is able to reproduce the magnitude and shape of
the observed multiplicity distributions quite well. The 
dashed lines are the results of
a multiphase transport model (AMPT)~\cite{zhang01,lin01}. This 
is a cascade model based on HIJING~\cite{wang91}, but includes 
final-state rescattering of produced particles. The AMPT model is also able
to account for the general trend of the measured distributions,
particularly for the most central collisions. We also plot the similar
distributions~\cite{Alner86} from $p\bar p$ collisions at $\sqrt{s}$=200~GeV, 
scaled up by the corresponding number of Au+Au participant pairs, for the
0-5\% and 40-50\% centralities. 
For central collisions the Au+Au data  
show a strong enhancement over the entire pseudorapidity 
range relative to the $p\bar p$ results, with
an excess of $(41\pm 9)\%$ observed at mid-rapidity. This suggests 
significant influence of the extended, high-density medium in the case 
of the heavy-ion collision. 
The observed enhancement decreases to 
about 10\% for 40-50\% centrality collisions.  
We  also note that the width in pseudorapidity of 
the measured distributions increases slightly as the  
centrality decreases, 
with $\sigma_{RMS}=2.33\pm 0.02$ and $2.40 \pm 0.02$ 
for the 0-5\% and 40-50\% centralities, respectively. 
This again suggests increased particle production at mid-rapidity
for more central collisions. These values can be compared to 
$\sigma_{RMS}= 2.38 \pm 0.05$  for the $p\bar p$ data.  
       
\begin{figure}
  \epsfig{file=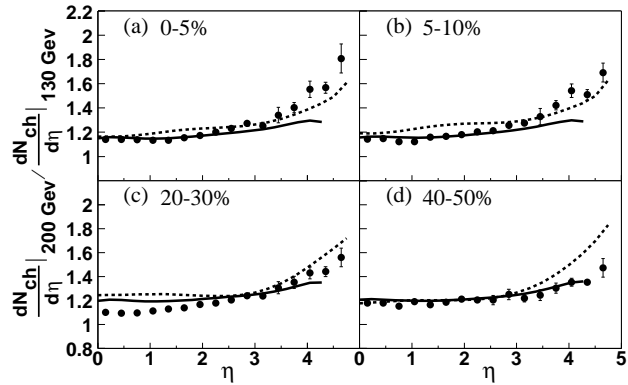,width=8.5cm}
  \caption{
    Ratio of $dN_{ch}/d\eta$ values 
    at $\sqrt{s_{NN}}$=200~GeV and 130~GeV
    compared to the model calculations (see Fig. 3 caption). 
    Total uncertainties are shown, assuming a 3\% relative scaling
    uncertainty between the two energies.}
  \label{dndeta_ratios}
\end{figure}

The ratios of  $dN_{ch}/d\eta$ 
values  measured at
$\sqrt{s_{NN}}$=130~GeV and $\sqrt{s_{NN}}$=200~GeV for different
centralities are shown in Fig.~\ref{dndeta_ratios}. An
increase in charged-particle multiplicity as a function of
energy for a central-plateau region ($|\eta| < 2.5$) is observed, 
with  a comparable increase  of 10\% to
20\% observed for all centralities. 
The upturn in the ratios seen at forward rapidities is due to the 
widening of the multiplicity distribution at the higher energy, 
consistent with the increase in beam rapidity ($\Delta y = 0.43$). 
The curves show the corresponding ratios resulting from the 
two model calculations.

\begin{figure}
  \epsfig{file=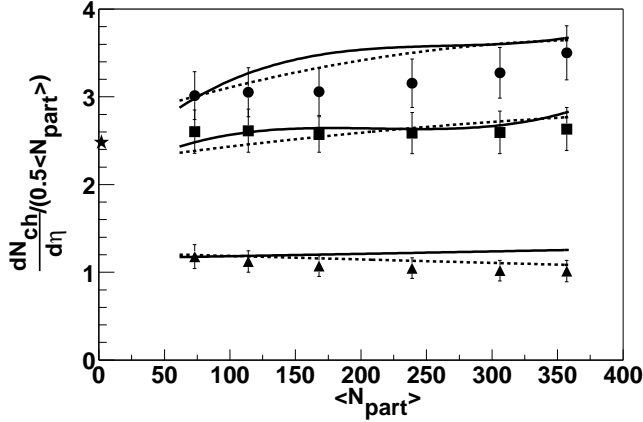,width=8.5cm}
  \caption{
    $dN_{ch}/d\eta$ per participant nucleon pair as a
    function of the average number of participants (see table)
    for $\eta$= 0 (circles), 3.0 (squares) and 4.5 (triangles). 
    The curves show the model predictions (see Fig. 3 caption). 
    The star denotes the $p\bar p$ result at 
    $\eta=0$~\cite{Alner86}.}
  \label{Npart}
\end{figure}

Finally in Fig.~\ref{Npart} we plot 
$(dN_{ch}/d\eta)/\langle N_{part}/2\rangle$
as a function of
the average number of participants $\langle N_{part}\rangle$ 
for three narrow pseudorapidity
regions ($\Delta \eta \approx 0.2$) around $\eta$ =0, 3.0 and
4.5. As already 
suggested, particle production per participant
pair is remarkably constant and near unity
at the  forward rapidities characteristic
of the fragmentation region, while showing a significant increase 
for the more central collisions with $\eta\approx 0$. 
The mid-rapidity behavior has been attributed to the
onset of hard-scattering processes which are dependent on the number of 
binary nucleon collisions
$N_{coll}$ rather than $N_{part}$~\cite{adcox01}. 
Using  $N_{coll}$ and $N_{part}$ values from HIJING
~\cite{wang91} we fit the data with a
function of the form 
$dN_{ch}/d\eta=\alpha\cdot N_{part}+\beta \cdot N_{coll}$. 
For $\eta=$ 0(4.5) we obtain: $\alpha=1.26 \pm 0.09 \pm 0.20$ 
($0.66 \pm 0.03 \pm 0.10$) 
and $\beta=0.15 \mp 0.04 \mp 0.05$($-0.06 \mp 0.01 \mp 0.03$), where the 
first uncertainty assumes a 3\%  point-to-point error
for the $dN_{ch}/d\eta$ values and
the second uncertainty results from the $N_{coll}$ 
and $N_{part}$
uncertainties.
For comparison we find $\alpha=1.24 \pm 0.08 \pm 0.20$
($0.55 \pm 0.02 \pm 0.09$) and 
$\beta=0.12 \mp 0.04 \mp 0.06$($-0.09 \mp 0.01 \mp 0.03$) at
$\sqrt{s_{NN}}$=130~GeV. For central events at $\eta\approx 0$ 
we find that the hard-scattering 
component to the charged-particle production remains
almost constant, with values of 
(20$\pm$7)\% and (25$\pm$7)\% at
$\sqrt{s_{NN}}$=130~GeV and 200~GeV, respectively. 
For this comparison, 
only the experimental component of the uncertainties 
are given since the
theory uncertainties will be highly correlated at the two
energies. 

In conclusion, we find that the charged-particle production 
increases by a constant amount from 
$\sqrt{s_{NN}}$=130~GeV to $\sqrt{s_{NN}}$=200~GeV in a
wide region around mid-rapidity.  
The data are well reproduced by
calculations based on high-density QCD and by the AMPT/HIJING
microscopic parton model. A phenomenological analysis in
terms of a superposition of soft- and hard-scattering 
particle production indicates that the  hard-scattering
component seen at  mid-rapidity for central collisions
is not significantly
enhanced as compared to 
$\sqrt{s_{NN}}$=130~GeV results. We find good
consistency with the gluon saturation model of Kharzeev and Levin, but
stress that within errors of models and data alike, the data can be
equally well reproduced by other models that do not require 
parton-collision saturation. This 
work establishes a baseline for particle production at the maximum
energy currently available for nucleus-nucleus collisions.

We thank the RHIC collider team for their efforts. 
This work was supported by the Division of Nuclear Physics
of the U.S. Department of Energy,  
the Danish Natural Science Research Council, the Research Council of
Norway, the Polish State Committee for Scientific Research (KBN) 
and the Romanian
Ministry of Research. 
We are grateful to D. Kharzeev, 
E. Levin, 
Zi-wei Lin, and 
H. Heiselberg for useful discussions and 
model calculations. 

\ifUseBibTeX 
\bibliography{dndeta}

\begin{thebibliography}{99}
\bibitem{sionly} We have reanalyzed our
  earlier $\sqrt{s_{NN}}$=130 GeV data to only include the same set of
  detectors as used here.
\end{thebibliography}
\else 

\fi 

\end{document}